\documentclass[%
 reprint,
 amsmath,amssymb,
 aps,prb
]{revtex4-1}

\usepackage{graphicx}% Include figure files
\usepackage{dcolumn}% Align table columns on decimal point
\usepackage{bm}% bold math
\usepackage{relsize}
\usepackage{braket}
\usepackage{amsmath}
\usepackage{color}
\usepackage{comment}

\begin{document}

%\preprint{APS/123-QED}

\title{Berry curvature for coupled waves of magnons and electromagnetic waves}% Force line breaks with \\

\author{Akihiro Okamoto$^1$, Ryuichi Shindou$^{2,3}$, and Shuichi Murakami$^1$}
\affiliation{%
 ${}^1$Department of Physics, Tokyo Institute of Technology, 2-12-1 Ookayama, Meguro-ku, Tokyo 152-8551, Japan\\
 ${}^2$International Center for Quantum Materials, School of Physics, Peking University, Beijing 100871, China\\
${}^3$Collaborative Innovation Center of Quantum Matter, Beijing 100871, China
}%% \altaffiliation[Also at ]{Physics Department, XYZ University.}%Lines break automatically or can be forced with \\
% \collaboration{MUSO Collaboration}%\noaffiliation

% \affiliation{
%  %This line break forced% with \\
% }%
% \affiliation{
%  Third institution, the second for Charlie Author
% \affiliation{%
%  Authors' institution and/or address\\
%  This line break forced with \textbackslash\textbackslash
% }%

\date{\today}% It is always \today, today,

\begin{abstract}
In this paper, we introduce Berry curvature, topological Chern number and topological chiral edge 
mode, that emerge from a hybridization between magnon and electromagnetic wave in a 
ferromagnet insulator. By focusing on the energy conservation, we first 
reformulate the Landau-Lifshitz-Maxwell equation into a Hermitian 
eigenvalue equation. From the eigenvalue 
equation, we define the Berry curvature of the magnon-photon coupled waves. 
We show that the Berry curvature thus introduced shows a prominent peak around a 
hybridization point between magnon mode and photon mode, and a massive hybrid   
mode takes a non-zero Chern number ($\pm 1$) due to the magnon-photon coupling. 
In accordance with the non-zero Chern number, the 
topological edge modes emerge inside the hybridization gap at a 
domain wall between two ferromagnetic insulators with opposite magnetizations.
\end{abstract}

\maketitle

%\tableofcontents

\section{\label{sec:introduction}Introduction}
Coupled waves between ferromagnetic moments and electromagnetic waves have been studied for a 
long time. Dispersion relations of the coupled waves of the magnon and the electromagnetic wave in layered film structures 
consisting of magnetic, ferroelectric, and insulating layers, are studied theoretically \cite{Demidov2000,doi:10.1063/1.1475373,Demidov2002} and experimentally. In recent years, coupling between quantum spins 
and photons have attracted much attention both in theory and in experiment. The coupled 
wave of spins and photons behaves differently depending on the strength of the coupling. When the 
coupling is strong, the wave is called a magnon-polariton \cite{0034-4885-37-7-001,LEHMEYER198532}. 
The magnon-polariton is promising for applications in quantum information 
science and technology. Recently, strong coupling between the Kittel mode and the cavity mode is 
studied in the YIG sphere \cite{PhysRevLett.113.156401,PhysRevLett.113.083603,10.1126/science.aaa3693}, 
film \cite{PhysRevB.91.094423}, and film split rings \cite{doi:10.1063/1.4904857}.
 
 The Berry curvature in various physical systems has also been attracting many researchers. A 
geometric character of the Bloch wavefunction gives rise to new phenomena such as topological electric and 
thermal Hall effect~\cite{RevModPhys.82.1959}. The Berry curvature have been studied in 
electrons\cite{PhysRevB.53.7010,PhysRevB.59.14915}, 
photons\cite{PhysRevE.70.026605,PhysRevLett.93.083901,PhysRevE.70.026605,PhysRevE.70.026605,PhysRevLett.96.073903,PhysRevE.74.066610,PhysRevLett.100.013904,PhysRevA.78.033834}, 
magnons\cite{Onose297,PhysRevB.84.184406,PhysRevLett.106.197202,PhysRevB.96.174437,PhysRevB.87.144101,PhysRevB.87.174402,PhysRevB.87.174427}, and so forth. 
Recently, calculations of finite Berry curvature are reported in various coupled systems such as 
systems with charge density and current coupling\cite{13486,jind}, exciton-photon 
coupling\cite{PhysRevX.5.031001}, and magnon-phonon coupling~\cite{PhysRevB.99.174435,shuzhang,gyungchoongo,sungjoonpark,pengtaoshen}. The hybridizations 
among these degrees of freedom lead to topological bands and novel edge states inside a 
hybridization gap. In the previous work, we have calculated the Berry curvature of magnetoelastic 
wave, by formulating a Hermitian eigenvalue equation from an equation of motion for the 
magnetoelastic wave\cite{PhysRevB.101.064424}. 

In this paper,
we formulate a Hermitian eigenvalue equation for coupled equations of motion for ferromagnetic moments
and electromagnetic waves. Based on the formulation, we calculate 
the Berry curvature of the coupled waves of magnons and 
electromagnetic waves\cite{label8551}. We find that the Berry curvature is prominently enhanced at a 
crossing point of the dispersions and we carify its asymptotic behavior around the crossing point. 
We wind that in the presence of the finite hybridization, the topological Chern number 
of the coupled wave becomes quantized to be non-zero integer. We show 
that in accordance with the non-zero Chern number, non-trivial topological edge modes of 
the coupled wave appear inside the hybridization gap at a domain wall. 

This paper is organized as follows. In Section  \ref{sec:formulation of eigenvalue equation}, we formulate 
generalized Hermitian eigenvalue equations from the equations of motion of magnons and 
electromagnetic waves and calculate eigenfrequencies. In Sections \ref{sec:berry curvature} 
and \ref{sec:edgemode}, we calculate the Berry curvature, the Chern number and its edge 
modes of the magnon and electromagnetic waves. We summarize the paper in Sec.~\ref{sec:conclusion}.

%This sample document demonstrates proper use of REV\TeX~4.1 (and
%\LaTeXe) in mansucripts prepared for submission to APS
%journals. Further information can be found in the REV\TeX~4.1
%documentation included in the distribution or available at
%\url{http://authors.aps.org/revtex4/}.

%When commands are referred to in this example file, they are always
%shown with their required arguments, using normal \TeX{} format. In
%this format, \verb+#1+, \verb+#2+, etc. stand for required
%author-supplied arguments to commands. For example, in
%\verb+\section{#1}+ the \verb+#1+ stands for the title text of the
%author's section heading, and in \verb+\title{#1}+ the \verb+#1+
%stands for the title text of the paper.

%Line breaks in section headings at all levels can be introduced using
%\textbackslash\textbackslash. A blank input line tells \TeX\ that the
%paragraph has ended. Note that top-level section headings are
%automatically uppercased. If a specific letter or word should appear in
%lowercase instead, you must escape it using \verb+\lowercase{#1}+ as
%in the word ``via'' above.s

\section{\label{sec:formulation of eigenvalue equation}Formulation of eigenvalue equation}
\begin{figure}[t]
 \begin{center}
 \includegraphics[height=2.5cm]{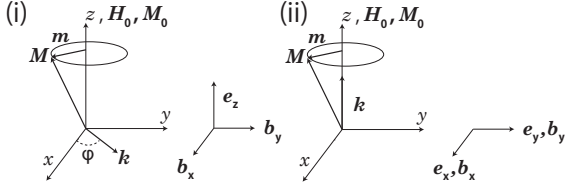}
 \end{center}
 \caption{\label{fig:geometry_EMW} Schematic illustration of the magnons and the electromagnetic waves for (i) the wavevector perpendicular to the saturation magnetization ($\theta=\pi/2$) and (ii) the wavevector parallel to the saturation magnetization $\theta=0$ (in Appendix A).}
\end{figure}

We consider a three-dimensional ferromagnetic insulator with isotropic electric permittivity. The saturation 
magnetization ${\bm M_0}$ and the applied magnetic field ${\bm H_0}$ are parallel to each other,  
and they are along the $z$-direction. The magnon field is described by a magnetization ${\bm m}$ 
in the $xy$ plane (Fig.~\ref{fig:geometry_EMW}). 
We assume that electromagnetic waves with the magnetic flux density ${\bm b}$ and the 
electric field ${\bm e}$ trasmit entirely through the ferromagnetic insulator without dissipation. 
The amplitudes of the magnon and the electromagnetic waves are proportional to 
$\exp{i({\bm k \cdot r}-\omega t)}$, with frequency $\omega$ and wavevector ${\bm k} 
\equiv (k_x,k_y,k_z) \equiv k (\sin \theta \cos\varphi,\sin\theta \sin \varphi,\cos\theta)$ 
(Fig.~\ref{fig:geometry_EMW}). Therefore, $\theta$ represents an angle between the wavevector 
and the saturation magnetization.  
The coupled equations of motions (EOM) consist of the Landau-Lifshitz equation and the 
Maxwell equation. 
In terms of the wavevector ${\bm k}$, the EOM take forms of 
 %are written as 
\begin{eqnarray}
\frac{ \partial {m_x}}{\partial t}&=&\frac{\omega_M}{4\pi}{b_y}-\omega_s{m_y}, \label{eq:EOM_mem_1} \\ 
\frac{ \partial {m_y}}{\partial t}&=&-\frac{\omega_M}{4\pi}{b_x}+\omega_s{m_x}, \\
\frac{ \partial {\bm{b}}}{\partial t}&=&-icK{\bm{e}}, \label{eq:EOM_mem_3}\\
\frac{ \partial {\bm{e}}}{\partial t}&=&i\frac{cK}{\epsilon}{(\bm{b}-4\pi\bm{m})}, \label{eq:EOM_mem_4}
\end{eqnarray}
where $\omega_s \equiv \omega_M+\omega_H$, $\omega_M \equiv 
4\pi g M_0$, $\omega_H \equiv g H_0$, $g$ is gyromagnetic constant, $c$ is the speed of light, 
$\epsilon$ is the permittivity, and ${K}$ is an anti-symmetric matrix defined as
\begin{eqnarray}
K= \left(\begin{array}{ccc}
     0 & -k_z & k_y \\
     k_z& 0 & -k_x \\
     -k_y & k_x & 0
    \end{array}
    \right).
\end{eqnarray}
For the later convenience, let us express 
Eqs.~(\ref{eq:EOM_mem_1})-(\ref{eq:EOM_mem_4}) as
\begin{eqnarray}
i\frac{\partial }{\partial t}{\bm x}_{\bm k}&=&H_{\rm eff}{\bm x}_{\bm k}, \label{eigenvalue_equation}
\end{eqnarray}
where ${\bm x}_{\bm k}$ is the eigenvector, 
${\bm x}_{\bm k}= {}^t({ m}_{\bm k,x},{ m}_{\bm k,y},\ {\bm b}_{\bm k},\ {\bm e}_{\bm k})$. The 8 $\times$ 8 matrix $H_{\rm eff}$ is given by
\begin{eqnarray}
H_{\rm eff}&=&\begin{pmatrix}
\omega_s{ \sigma_2} & -\frac{\omega_M}{4\pi}{ \sigma'_2} & { 0} \\
{ 0} & { 0} & cK\ \\
-\frac{4\pi c}{\epsilon }(K')^t\ & -\frac{c}{\epsilon }K\ &{ 0}  \\
\end{pmatrix}
\end{eqnarray}
%where we introduced matrices 
with a 2 $\times$ 2 matrix ${ \sigma_2}$ and 2 $\times$ 3 matrices $I',\  { \sigma'_2},\ {K'}$:
\begin{eqnarray}
{I'}&=&
 \left(\begin{array}{ccc}
     1 & 0 & 0 \\
     0 & 1 & 0 \\
    \end{array}
    \right), \\
{ \sigma_2}&=&
 \left(\begin{array}{ccc}
     0 & -i \\
     i& 0 \\
    \end{array}
    \right), \ \ { \sigma'_2}=
 \left(\begin{array}{ccc}
     0 & -i & 0 \\
     i& 0 & 0 \\
    \end{array}
    \right), \\
 {K'}&=&
 \left(\begin{array}{ccc}
      0 & -k_z & k_y \\
     k_z & 0 & -k_x \\
    \end{array}
    \right).
\end{eqnarray}
We call $H_{\rm eff}$ an effective Hamiltonian. 

To define the Berry curvature for the coupled wave from the EOM, 
let us assume that a constant Hermitian matrix $\gamma$ makes $H_{\rm eff}$ to be Hermtian as 
$\tilde{H}_{\rm eff} \equiv  \gamma H_{\rm eff}=\tilde{H}^{\dagger}_{\rm eff}$.
In terms of these Hermitian matrices, the coupled EOM reduces to 
\begin{eqnarray}
i \gamma\frac{\partial {\bm x}_{\bm k}}{\partial t}&=&\tilde{H}_{\rm eff}{\bm x}_{\bm k}. \label{eigenvalue_equation_Hermite}
\end{eqnarray}
Define a `norm' of ${\bm x}_{\bm k}$ in terms of the Hermitian matrix $\gamma$ as
${\bm x}_{\bm k}^\dagger \gamma {\bm x}_{\bm k}$. Since 
$\gamma$ and $\tilde{H}_{\rm eff}$ are both Hermitian, one see that the norm is a constant 
of motion, $\partial ({\bm x}_{\bm k}^\dagger \gamma {\bm x}_{\bm k})/\partial t=0$. Physically 
speaking, the constant of motion must correspond to a total energy density of the system. 
Thus, we choose the Hermtian matrix $\gamma$ as 
\begin{eqnarray}
\gamma&=& \begin{pmatrix}
     \frac{(4\pi)^2\omega_s}{\omega_M}I & -4\pi I' &  {{0}}  \\
     -4\pi (I')^t & I &  {{0}} \\
      {{0}}  &  {{0}}  & \epsilon I 
    \end{pmatrix}, 
\end{eqnarray} 
with  
\begin{align}
\tilde{H}_{\rm eff}(\bm{k})=&\gamma H_{\rm eff}(\bm{k}) \nonumber \\
=&\begin{pmatrix}
\frac{{(4\pi\omega_s)}^2}{\omega_M}{\sigma_2} & -{4\pi}\omega_s{ \sigma'_2}&  -4\pi cK'\ \\
-{4\pi}\omega_s({ \sigma'_2})^\dagger & {\omega_M}{\Sigma_2} & cK & \\
-{4\pi c}(K')^t & -{c}K & {\bm 0}   \\
\end{pmatrix}, \nonumber \\
\end{align}
and a 3 by 3 matrix $\Sigma_2$, 
\begin{eqnarray}
 { \Sigma_2}&=&
 \left(\begin{array}{ccc}
     0 & -i & 0 \\
     i& 0 & 0 \\
     0& 0 & 0 \\
    \end{array}
    \right).
\end{eqnarray} 
Then, the norm is equal to the total energy density, consisting of the energy density of the electric wave 
$u_{e}$ and that of the the magnetic wave 
$\ u_{m}$~\cite{1067273,1133103,doi:10.1063/1.332451}, 
\begin{align}
{\bm x}_{\bm k}^\dagger \gamma {\bm x}_{\bm k} &= u_e + u_m, \label{ut} \\
u_{e}&={\bm e}_{\bm k}^\dagger \frac{\partial (\omega \hat {\epsilon })}{\partial \omega}{\bm e}_{\bm k}=\epsilon|{\bm e}_{\bm k}|^2, \label{ued} \\ 
u_{m}&={\bm h}_{\bm k}^\dagger  \frac{\partial (\omega \hat {\mu})}{\partial \omega}{\bm h}_{\bm k}= (4\pi )^2\frac{\omega_H}{\omega_M}|{\bm m}_{\bm k}|^2+|{\bm h}_{\bm k}|^2.  \label{umd}
\end{align}
Here $\hat {\epsilon}$ and $ \hat {\mu}$ are an isotropic permittive tensor and permeability 
tensor defined by ${\bm b}_{\bm k}=\hat{\mu} {\bm h}_{\bm k}$ where ${\bm h}_{\bm k}$ represents a magnetic field. We henceforth choose a normalization 
condition of the eigenvector ${\bm x}_{\bm k}$ 
as ${\bm x}_{\bm k}^\dagger \gamma {\bm x}_{\bm k}=1$. 

The eigenvalue equation (\ref{eigenvalue_equation_Hermite}) gives an equation for the dispersion relation
\begin{widetext}
\begin{eqnarray}
\omega^6 -\left(2\omega_{em}^2+\omega_s^2 \right)\omega^4 \nonumber+\omega_{em}^2\left(\omega_{em}^2+2\omega_H \omega_s+\omega_M \omega_s \sin^2{\theta }\right)\omega^2\nonumber-\omega_H\omega_{em}^4\left(\omega_H+\omega_M\sin^2{\theta}\right)=0, \nonumber \\ \label{dispersion relation}
\end{eqnarray} 
\end{widetext}
where $\omega_{em}=ck/\sqrt{\epsilon }$. The dispersion relation in Eq.~(\ref{dispersion relation}) has only six solutions, while the dimension of the eigenvalue equation (\ref{eigenvalue_equation}) is eight. 
The other two are nothing but two zero modes that correspond to unphysical gauge degrees of 
freedom. Namely, Eqs.~(\ref{eq:EOM_mem_3}) and (\ref{eq:EOM_mem_4}) satisfy 
${\bm k}\cdot {\bm b}=0 $ and 
${\bm k}\cdot {\bm e}=0 $, respectively and correspondingly, Eq.~(\ref{eigenvalue_equation}) always has 
two eigenvectors that belong to the zero eigenfrequency. 
The six physical solutions consist of pairs of positive and negative frequencies.
In the following, we only consider the case 
of (i) $\theta=\pi/2$ as shown in Fig.~\ref{fig:geometry_EMW}(i). 
We leave the case of (ii) $\theta=0$ in Appendix A.

\section{\label{sec:berry curvature}Berry curvature of coupled waves between magnons and electromagnetic waves 
for the case with $\theta=\pi/2$ ($k_z=0$)}
%Here we consider the case with . 
When ${\bm k}$ is perpendicular to the magnetization ${ M}_0$ 
($\theta=\pi/2$), the wavevector becomes the two-dimensional 
vector, ${\bm k}=(k_x,k_y) = k(\cos\varphi,\sin\varphi)$, and 
Eq.~(\ref{dispersion relation}) reduces to
\begin{eqnarray}
\left[\omega ^4-(\omega_{em}^2+\omega_s^2)\omega^2+\omega_{em}^2\omega_s\omega_H \right]\left[\omega^2-\omega_{em}^2 \right]=0. \nonumber \\ \label{kperpM}
\end{eqnarray} 
The eigenfrequencies from the first and second parentheses correspond to the set of 
components $m_x, m_y, b_x, b_y, e_z$, and that of $b_z, e_x, e_y$, respectively. 
The decoupling between these two sets is due to a mirror symmetry with respect to 
the $xy$ plane, under which the wavevector ${\bm k}$ is invariant for the case with 
$\theta=\pi/2$.  The first set comprises the hybrid waves of a magnon and an electromagnetic wave, 
\begin{eqnarray}
\omega^2_{\pm}=\frac{\omega_s^2+\omega_{em}^2}{2}\pm\sqrt{\left(\frac{\omega_s^2-\omega_{em}^2}{2}\right)^2+(\zeta k)^2}. \label{Dis_hyb}
\end{eqnarray}
Meanwhile within the linearized EOM, 
the second set of the fields ($b_z, e_x, e_y$) represents a pure and is free from the hybridization with magnon with its frequency equal to $\omega_{em}$, satisfying $\omega_{-} <\omega_{em} <\omega_{+}$ (Figs.~\ref{fig:BC_weak_EMW}(a) and ~\ref{fig:BC_strong_EMW}(a)). Here 
\begin{eqnarray}
\zeta \equiv \sqrt{\frac{\omega_M\omega_s}{\epsilon}}c  \label{zeta}
\end{eqnarray}
stands for the hybridization strength 
between magnon and electromagnetic waves. For the $\omega_{+}$ branch of 
Eq.~(\ref{Dis_hyb}), the dispersion at $k\rightarrow \infty $ 
and $k\rightarrow  0$ has the following asymptotic forms, 
\begin{eqnarray}
\omega_+^2 \simeq
\begin{cases}
\omega^2_{em}+\frac{(\zeta k)^2}{\omega^2_{em}} & (k\rightarrow \infty ), \\
    \omega^2_{s}+\frac{(\zeta k)^2}{\omega^2_{s}} & (k\rightarrow 0),
  \end{cases} \label{dispersion_+}
\end{eqnarray}
and for the $\omega_{-}$ branch of Eq.~(\ref{Dis_hyb}) is 
\begin{eqnarray}
\omega_- ^2 \simeq
\begin{cases}
    \omega^2_s-\frac{(\zeta k)^2}{\omega^2_{em}}=\omega^2_{mag} & 
(k\rightarrow \infty ), \\
    \omega^2_{em}-\frac{(\zeta k)^2}{\omega^2_{s}} & 
(k\rightarrow 0),
  \end{cases} \label{dispersion_-}
\end{eqnarray}
%At $k\rightarrow \infty $, the dispersion of the $\omega_{-} \equiv \omega_1$ branch
%is also written as
%\begin{eqnarray}
%\omega_{1}^2 = \omega^2_{mag}-\frac{\omega^4_s}{4\omega^2_{em}}. \label{dispersion_theta_90_k_s_1}
%\end{eqnarray}
where $\omega^2_{mag}=\omega_H\omega_s=\omega_H(\omega_H+\omega_M) $ is the dispersion of the magnon in the magnetostatic regime.

Let $k^*$ and $\tilde{\omega}$ denote the wavenumber and frequency 
at a crossing point between the dispersions of the magnon $\omega_s$ and the electromagnetic 
wave $\omega_{em}$ without the coupling ($\zeta=0$);  
\begin{eqnarray}
\omega_s=\omega_{em}(k^*) \equiv 
\tilde{\omega},  \ \ k^* \equiv \frac{\sqrt{\epsilon} \omega_s}{c}.  \label{HG0}
\end{eqnarray}
The frequencies of the coupled wave at the crossing point $k=k^*$ is 
given by  
\begin{eqnarray}
\omega_{\pm} =\sqrt{\tilde{\omega}^2\pm \tilde{\omega} \!\ \Delta \omega}, 
\ \ \Delta \omega \equiv \frac{\zeta k^*}{\tilde{\omega}}. \label{HG}
\end{eqnarray}
where $\Delta \omega$ is defined as a hybridization gap at the crossing point.
Note that the crossing point is located outside the magnetostatic regime. 
%When the magnon satisfies $\nabla \times {\bm h}_{\bm k}\approx {\bm 0}$, it is called magneto static wave. 
By using Eqs.~(\ref{eq:EOM_mem_3})-(\ref{eq:EOM_mem_4}), the magnetic field and the 
magnetization are written as\cite{stancil}
\begin{align}
&{\bm h}_{\bm k}=\frac{4\pi}{1-\omega^2/\omega_{em}^2}\left(-\frac{{\bm k}\cdot{\bm m}_{\bm k}}{k^2}{\bm k} +\frac{\omega^2}{\omega_{em}^2}{\bm m}_{\bm k}\right), \\ 
&{\bm k} \times {\bm h}_{\bm k}=\frac{\omega^2/\omega_{em}^2}{1-\omega^2/\omega_{em}^2}{\bm k}\times{\bm m}_{\bm k}. 
\end{align} 
The magnetostatic regime is defined by $\omega \ll \omega_{em}$, where the magnetic field becomes
approximately rotation free, ${\bm h}_{\bm k}\simeq -4\pi({{\bm k}\cdot{\bm m}_{\bm k}}){\bm k}/{k^2}$ and $
{\bm k} \times {\bm h}_{\bm k}\simeq {\bm 0}$. It is obvious that the crossing point 
($\omega \simeq \omega_{em}$) sits far outside the magnetostatic regime. In the following, we will show 
that the Berry curvature of the coupled modes shows a prominent peak near the crossing point outside the magnetostatic regime. 

%Next we focus on the eigenmodes $\omega_{\pm}$, giving 
The coupled modes between magnons and electromagnetic waves involve the 
components $m_x, m_y, b_x, b_y$, and $e_z$. The eigenvalue equation for the coupled modes 
is given by a 5 $\times$ 5 matrix extracted from $\tilde{H}_{\rm eff}$: 
\begin{eqnarray}
\tilde{H}^{\perp}_{\rm eff}(\bm{k}){\bm x}_{{\bm k},\pm}&=
&\omega_{\pm} \gamma^{\perp} {\bm x}_{{\bm k},\pm}, \label{EV_Eq}
\end{eqnarray} 
with
\begin{eqnarray}
\tilde{H}^{\perp}_{\rm eff}(\bm{k})&=&\scalebox{1.0}{$
\begin{pmatrix}
     0 & -i\frac{(4\pi)^2\omega_s^2}{\omega_M} & 0 & 4\pi i\omega_s & -4\pi ck_y \\
     i\frac{(4\pi)^2\omega_s^2}{\omega_M} & 0 & -4\pi i\omega_s & 0 & 4\pi ck_x \\
     0 & 4\pi i\omega_s & 0 & -i\omega_M & ck_y \\
     -4\pi i\omega_s & 0 & i\omega_M & 0 & -ck_x \\
     -4\pi ck_y & 4\pi ck_x & ck_y & -ck_x & 0 \\
    \end{pmatrix} 
$}. \nonumber \\  \label{Hamiltonian1}
\end{eqnarray} 
The energy density of the hybridized modes is given by a norm of a five-components
eigenvector, ${\bm x}_{\bm k} \equiv 
{}^t(m_{{\bm k},x},m_{{\bm k},y}, b_{{\bm k},x} , b_{{\bm k},y},e_{{\bm k},z})$. 
The norm is given by ${\bm x}^\dagger_{\bm k} \gamma^{\perp} 
{\bm x}_{\bm k}$, where $\gamma^{\perp}$ is  a 5 $\times$ 5 hermitian matrix extracted from the matrix $\gamma$:  
with  
\begin{eqnarray}
&&\gamma^{\perp} \equiv \begin{pmatrix}
     \frac{(4\pi)^2\omega_s}{\omega_M} & 0 & -4\pi & 0 & 0 \\
     0 & \frac{(4\pi)^2\omega_s}{\omega_M} & 0 & -4\pi & 0 \\
     -4\pi & 0 & 1 & 0 & 0 \\
     0 & -4\pi & 0 & 1 & 0 \\
     0 & 0 & 0 & 0 & \epsilon \\
    \end{pmatrix}. 
\end{eqnarray}
Based on this normalization, the Berry curvature of the coupled modes for the 
$\omega_{\pm}$ branches  
 is defined as
\begin{eqnarray}
\Omega _{z, n}(\bm{k})&=&i\epsilon _{\alpha \beta}\frac{\partial \bm{x}^\dagger_{\bm{k},n}}{\partial k_\alpha }\gamma^{\perp}\frac{\partial \bm{x}_{\bm{k},n}}{\partial k_\beta  }
\label{berrycurvature}
\end{eqnarray}
for $n=\pm$, ${\bm k} = (k_x,k_y)=k(\cos\varphi,\sin\varphi)$, and $\alpha,\beta=x,y$ 
with ${\bm{x}}^\dagger_{{\bm k},n}\gamma^{\perp}{\bm{x}}_{{\bm k},n}=1$. 
Here $\epsilon _{\alpha \beta}$ is the antisymmetric tensor with $\epsilon _{xy}=-\epsilon _{yx}=1$.
After a lengthy calculation, we find that the Berry curvature depends only on $k$; 
\begin{eqnarray}
\Omega_{z,\pm}(k)=\frac{1}{k}\frac{\partial}{\partial k}\left(\frac{\frac{\omega_{\pm}}{\omega_s}(\omega_{\pm}^2-\omega_{em}^2)}{(2\omega_{\pm}^2-\omega_s^2-\omega_{em}^2)}\right). \label{BC_MEMW}
\end{eqnarray}
The details of the derivation are shown in Appendix B. By the similar procedure as in the magnetoelastic 
wave\cite{PhysRevB.101.064424}, we henceforth calculate the Berry curvature in the regimes with weak and strong coupling 
defined by $\Delta \omega/ \tilde{\omega}\ll 1$ and $\Delta \omega/ \tilde{\omega}\simeq  1$, respectively.

\subsection{Weak coupling regime}
\begin{figure}[t]
 \begin{center}
 \includegraphics[width=8.0cm]{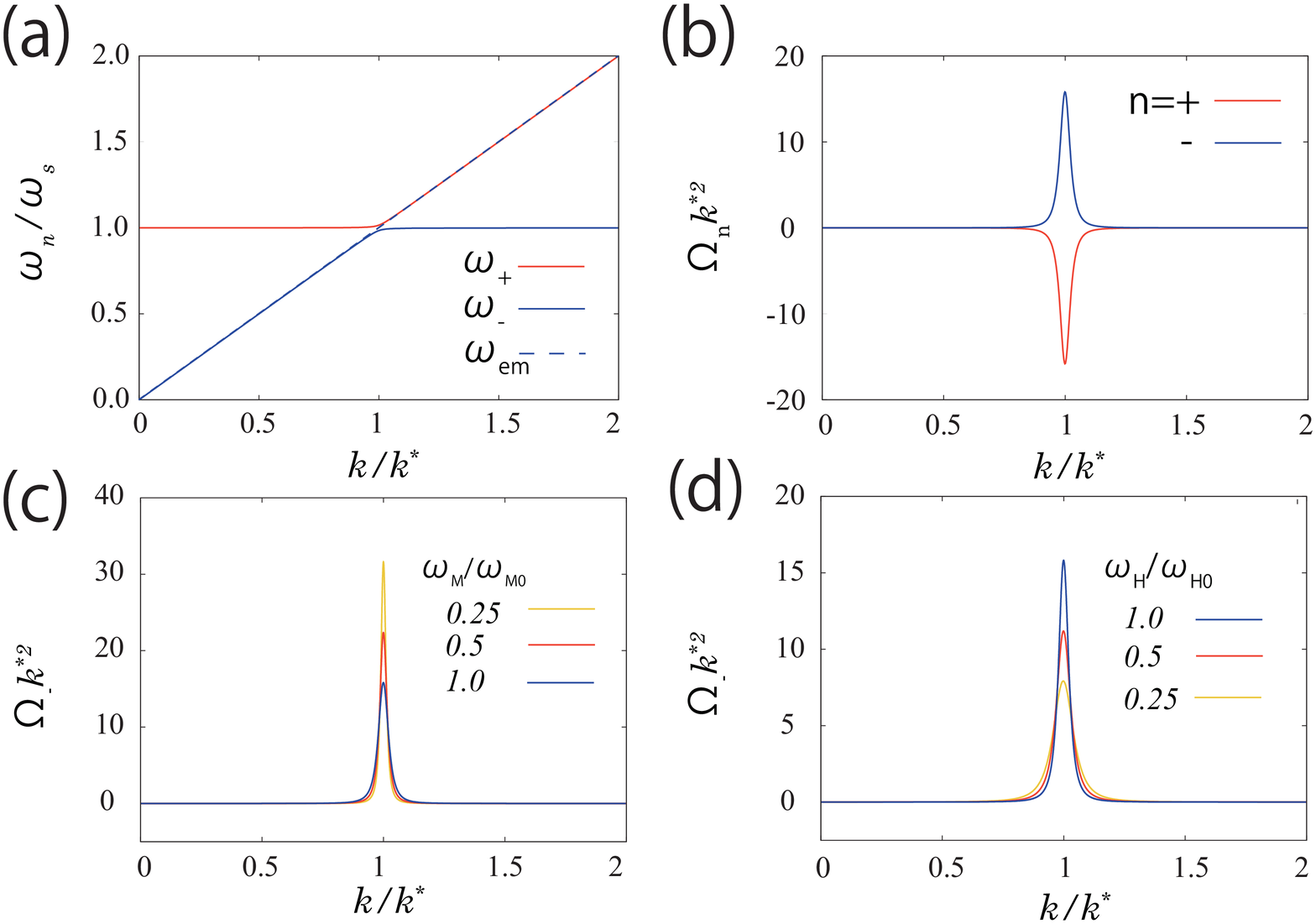}
 \end{center}
 \caption{\label{fig:BC_weak_EMW} (Color online) Dispersions and Berry curvatures for coupled modes between magnons and electromagnetic waves in a weak coupling regime. 
(a) Dispersions of the $\omega_{+}$ and $\omega_{-}$ branches, 
and (b) Berry curvatures of the $\omega_{+}$ and $\omega_{-}$ branches 
as functions of $k$. The parameters are set to be $\omega _H/\omega _M=10^3$ with $\omega_H=5$ [{\rm GHz}]. Berry curvatures for the $\omega_{-}$ mode is shown for (c) $\omega_M/\omega_{M0}=0.25, 0.5, 1.0$ with $\omega_H/\omega_{H0}=1.0$, and (d) $\omega_H/\omega_{H0}=\ 0.25,\ 0.5,\ 1.0$ with $\omega_M/\omega_{M0}=1.0$, with 
$\omega_{M0}=5\times 10^{-3}$ [{\rm GHz}] and $\omega_{H0}=5$ [{\rm GHz}].
 }
\end{figure}

The weak-coupling regime between magnon and electromagnetic wave is expressed 
as $\omega_M\ll \omega_s$ from Eqs.~(\ref{zeta}), (\ref{HG0}), and (\ref{HG}). 
To satisfy this condition, we set $\omega_M\ll \omega_H$ to calculate the Berry curvature. 
When $\omega_M\ll \omega_H$, the hybridization gap is approximately evaluate as
\begin{eqnarray}
\Delta \omega \simeq \sqrt{\omega_M \omega_H}.
\end{eqnarray} 
The gap is much smaller than $\tilde{\omega}$ under this condition. 
We show the results of the numerical calculation of the dispersion and the Berry curvature in Figs. \ref{fig:BC_weak_EMW} (a) and (b). The Berry curvatures for $\omega_{\pm}$ show a strong peak, and are localized at the crossing point of 
the dispersions.

The peak value of the Berry curvature at the 
crossing point ($k=k^{*}$) is approximately evaluated as
\begin{eqnarray}
\Omega_{z,\pm}(k=k^*)&=&\mp \frac{\tilde {\omega}^2}{2{k^*}^3\zeta}=\mp \frac{1}{{2k^*}^2\Delta \omega/\tilde{\omega}}. \label{BC_weak}
\end{eqnarray} 
In terms of $\tilde{\omega}\simeq \omega_H$, and 
$\Delta \omega \simeq \sqrt{\omega_M \omega_H}$, we can see that the Berry curvature is proportional to
$\Omega_{\pm}(k^*)\propto  {1}/{\omega_M ^{1/2}}$ and ${1}/{\omega_H ^{3/2}} $. The dependences of the Berry curvature on $\omega_M$ and $\omega_H$ agree with Figs.~\ref{fig:BC_weak_EMW} (c) and (d). This result has the same form as the result of the magnetoelastic wave with respect to the hybridization gap except for some coefficients\cite{PhysRevB.101.064424}. It means that the main effect of the Berry curvature induced by the hybridization 
has a universal feature around the hybridization gap in the weak coupling regime. 

\subsection{Strong coupling regime}
\begin{figure}[t]
 \begin{center}
 \includegraphics[width=8.0cm]{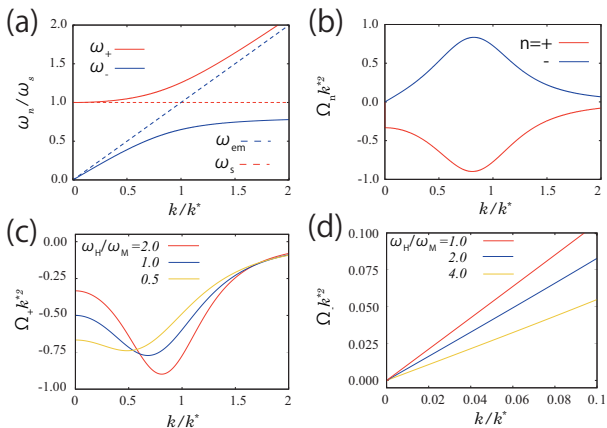}
 \end{center}
 \caption{\label{fig:BC_strong_EMW} (Color online) Dispersions and Berry curvatures for coupled modes between magnons and electromagnetic waves in strong coupling regime 
(a) Dispersions of the $\omega_{+}$ and $\omega_{-}$ branches. 
(b) Berry curvatures of the $\omega_{+}$ and $\omega_{-}$ branches as functions of 
$k$. $\omega _H/\omega _M=2.0$ with $\omega_H=5.0$ [{\rm GHz}]. 
The Berry curvature of the $\omega _+$ mode is shown for 
(c) $\omega_H/\omega_{M}=0.5, 1.0, 2.0$, and 
the $\omega _-$ mode for (d) $\omega_H/\omega_{M}=\ 0.5,\ 1/\sqrt{2},\ 1.0$.}
\end{figure}

To calculate the Berry curvature in the strong-coupling regime, we set $\omega_M\simeq \omega_s$.
The results of the dispersion and the Berry curvature are shown in Fig.~\ref{fig:BC_strong_EMW}.
When the coupling between magnon and electromagnetic wave is strong, the peak of the Berry 
curvature at $k=k^*$ broadens as shown in Figs.~\ref{fig:BC_strong_EMW} (a) and (b). 
 
The Berry curvature of the coupled wave is affected by the hybridization even at $k\ll k^*$.  
By using Eq.~(\ref{BC_MEMW}) and the dispersions Eqs.~(\ref{dispersion_+}) and 
(\ref{dispersion_-}), we obtain the 
Berry curvature for $k\ll k^*$
\begin{eqnarray}
\Omega_{z,+}(k)&\sim & -\frac{\zeta^2}{\omega_s^4}, \label{} \\
\Omega_{z,-}(k)&\sim &\frac{3\zeta^2c' k }{\omega_s^5}, \label{}
\end{eqnarray} 
where $c'=\sqrt{c^2/\epsilon-\zeta ^2/\omega_s^2}$. These results show that the Berry curvature 
for $k\ll k^*$ is strongly affected by the coupling $\zeta$. The Berry curvature $\Omega_{z,+}$ is 
finite at $k\rightarrow 0$, while the Berry curvature $\Omega_{z,-}$ is zero at $k\rightarrow 0$. 
The analytical results agree with the result of Figs.~\ref{fig:BC_strong_EMW} (c) and (d). 

The asymptotic behavior of $\Omega_{z,-}$ around $k \simeq 0$ comes from the  
linearly polarized nature of the magnetic field and flux in the vicinity of $k=0$.
For simplicity, we choose the wavevector ${\bm k}=k{\bm e}_y$ where ${\bm e}_y$ is a 
unit vector along the $y$-axis. A relation between $h_x$ and $h_y$ is written for the $\omega_{em}$ mode as\cite{stancil}
\begin{eqnarray}
\frac{h_y}{h_x}=-\frac{i \omega \omega_M}{\omega_{mag}^2 -\omega ^2}.
\end{eqnarray} 
Thus, the magnetic field becomes linearly polarized along the $x$ direction when $k\rightarrow 0$.
In addition, the magnetic flux also becomes linearly polarized along $x$ 
at $k\rightarrow 0$, because the non-diagonal component of the permeability 
tensor $\mu_{xy}$ becomes smaller at $k\rightarrow 0$. These behaviors are the same 
for an arbitrary direction of ${\bm k}$. Thus, the eigenvector becomes asymptotically independent of $k$ and 
the Berry curvature $\Omega_{z,-}(k)$ becomes zero at $k\rightarrow 0$. 
This asymptotic behavior of the Berry curvature for the $\omega=\omega_-$ mode is totally different from that of 
the magnetoelastic wave $\omega=\omega_+$ in the strong coupling, where the Berry 
curvature of the linearly dispersive branch diverges toward $k = 0$.  

\section{\label{sec:edgemode}Topological edge modes at $k_z=0$}
\subsection{Chern number}
\begin{figure}[t]
 \begin{center}
 \includegraphics[width=9.0cm]{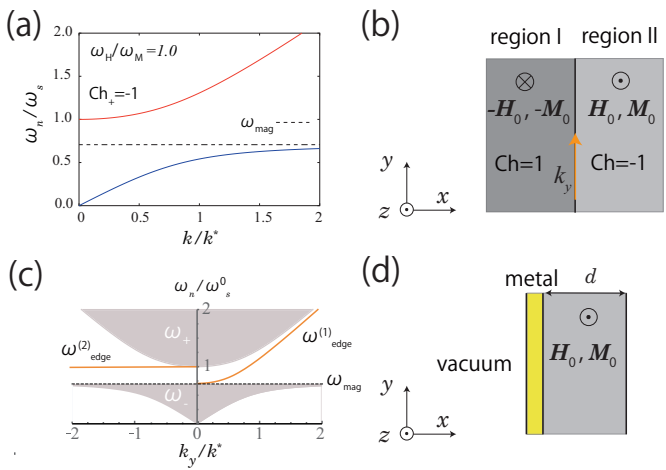}
 \end{center}
 \caption{\label{fig:Chern_Edge} (Color online) Topological Chern number and 
topological edge modes of coupled waves between magnons and electromagnetic waves (a) Dispersions of the $\omega_{+}$ and $\omega_{-}$ branches with the 
Chern number for the $\omega_+$ branch. (b) Schematic picture of two ferromagnetic regions with opposite magnetization 
and magnetic field. In region I, the Chern number of the $\omega_{+}$ branch is $+1$ with 
${\bm H}=(0,0,-H_0)$, and in region II it is $-1$ with 
${\bm H}=(0,0,H_0)$. The boundary between the two regions is parallel to the $y$-axis.  
We define the surface momentum along $y$ as $k_y$. (c) Dispersions for 
the edge modes (orange lines) with 
$\omega^0_H=\omega^0_M=5.0$ [{\rm GHz}]. Gray-colored regions 
show projections of the dispersions of the bulk modes $\omega_{+}$ and $\omega_{-}$ onto the surface momentum $k_y$. 
(d) Ferromagnetic insulator slab with a metalized surface. A chiral mode appears only in the $k_y<0$ on the metalized surface.}
\end{figure}

Let us define an integral of the Berry 
curvature for the $\omega_{n}$ branch over the two-dimensional momentum space~\cite{PhysRevLett.49.405,KOHMOTO1985343};  
\begin{eqnarray}
{\rm Ch}_n=\int^{+\infty}_{-\infty} 
\int^{+\infty}_{-\infty}\Omega_{z,n}(\bm k)  \!\ \frac{dk_x dk_y}{2\pi}, 
\end{eqnarray}
with $n=\pm$. The integral is quantized to be an integer (Chern number), when the 
$\omega_n$ branch is separated from the other branches by a direct gap for any 
${\bm k}=(k_x,k_y)$. The quantized integer is identical with a number of 
topological chiral edge modes inside the gap~\cite{PhysRevB.25.2185}. The edge modes are localized along 
a boundary of the system within the $xy$ plane.  
Using Eq.~(\ref{BC_MEMW}), we obtain 
\begin{eqnarray}
{\rm Ch}_n&=&\int^{\infty }_{0} d{ k}\frac{\partial}{\partial k}\left(\frac{\frac{\omega_{n}}{\omega_s}(\omega_{n}^2-\omega_{em}^2)}{(2\omega_{n}^2-\omega_s^2-\omega_{em}^2)}\right) \nonumber \\
&=&N_{n}(\infty )-N_{n}(0) ,
\end{eqnarray} 
where
\begin{eqnarray}
N_{n}(k_0)\equiv \left . \left(\frac{\frac{\omega_{n}}{\omega_s}(\omega_{n}^2-\omega_{em}^2)}{(2\omega_{n}^2-\omega_s^2-\omega_{em}^2)}\right) \right |_{k\rightarrow k_0}.
\end{eqnarray} 
Using Eqs.~(\ref{dispersion_+}) and (\ref{dispersion_-}), we have
\begin{eqnarray}
N_{+}(k) =\begin{cases}
    0 & (k\rightarrow \infty ) \\
    1 & (k\rightarrow 0)
  \end{cases},  
\end{eqnarray} 
and 
\begin{eqnarray}
N_{-}(k) =\begin{cases}
    \frac{\omega_{mag}}{\omega_s} & (k\rightarrow \infty) \\
    0 & (k\rightarrow 0)
  \end{cases}.  
\end{eqnarray} 
Thus, the Chern number for the $\omega_{+}$ branch is $-1$,
\begin{eqnarray}
{\rm Ch}_+=-1.
\end{eqnarray}
The dispersion and its Chern number are illustrated in Fig.~\ref{fig:Chern_Edge} (a). 
From the quantization of the Chern number, we expect that 
a chiral edge mode with $k_z=0$ appears inside the hybridization gap between the 
$\omega_+$ branch and $\omega_-$ branch.

The integral of the Berry curvature for the $\omega_{-}$ branch is 
not quantized to an integer. This is because the $\omega_-$ branch 
in the particle space ($\omega=\omega_{-}\ge 0$) and its hole counterpart
 ($\omega=-\omega_{-}\le 0$) forms a band touching at $k=0$; $\omega_{-}(k=0)=0$. 
In the eigenvalue equation (\ref{eigenvalue_equation}), the branch with the 
positive frequency and that with the negative frequency are coupled with each other. 
Due to the band touching at $k=0$, the Chern number for the $\omega_{-}$ branch is not well 
defined. As a result, the sum of the Chern number over the branches with the positive frequency 
region is not zero either, unlike the cases with a gap between the positive $\omega$ and the negative $\omega$ branches\cite{PhysRevB.87.174427}. 

\subsection{Chiral edge modes}
From the quantization of the Chern number of the $\omega_{+}$ branch, we expect that a chiral edge 
mode with $k_z=0$ appear inside the hybridization gap. The mode is localized at a boundary between topologically different regions. Here, we show an emergence of such topological chiral edge modes at an interface between two regions with opposite magnetizations. We consider a domain wall as schematically illustrated in Fig.~\ref{fig:Chern_Edge} (b). The magnetization and magnetic field 
is directed along $-z$ direction in region I ($x< 0$) and $+z$ direction in region II ($x> 0$). Namely, ${\bm H}=(0,0, H_0)$, ${\bm M}=(0,0, M_0)$ in region II and ${\bm H}=(0,0, -H_0)$, ${\bm M}=(0,0, -M_0)$ in region I, where $H_0$ and $M_0$ are positive. This means $\omega_H=\omega^0_H\equiv gH_0$, $\omega_M=\omega^0_M\equiv 4\pi gM_0$, and $\omega_s=\omega^0_s\equiv\omega^0_H+\omega^0_M$ in region II and $\omega_H=-\omega^0_H$, $\omega_M=-\omega^0_M$, and $\omega_s=-\omega^0_s$ in region I.  
From Eq.~(\ref{BC_MEMW}), 
the Berry curvature for the $\omega_{n}$ branch changes its sign from the 
region I to the region II. Thus, the Chern number for the $\omega_+$ branch in the region I is $+1$, 
while that in the  region II is $-1$.  
 
The number of chiral edge modes at an interface with two regions with different Chern numbers 
equals to the difference of the two Chern numbers between the two regions\cite{PhysRevB.25.2185}. It is independent of the details of the interface. Since the Chern 
number in the region I and in the region II are $1$ and $-1$ respectively, two chiral 
edge modes are expected to emerge at the interface. To see them, we note that 
the wavenumber $k_y$ along the edge ($y$ axis) is conserved, while we should replace $k_x$ by $-i\partial_x$ 
in Eq.~(\ref{Hamiltonian1}). 
We then calculate eigenmodes of Eq.~(\ref{EV_Eq}) at the boundary.   
The eigenmodes localized at the boundary is proportional to $e^{ik_y y +\kappa x}$ 
for $x<0$ and $e^{ik_y y -\kappa x}$ for $x>0$ with $\kappa>0$. From the 
boundary conditions for the electromagnetic waves $b_x$, $h_y$ and $e_z$, we obtain two 
edge-mode solutions $\omega=\omega^{(1)}_{\rm edge},\ \omega^{(2)}_{\rm edge}$ inside the gap between 
$\omega_+$ and $\omega_-$ (see Appendix E), and their dispersion relations are 
shown in Fig.~\ref{fig:Chern_Edge} (c). 
The dispersions of the edge modes are written as
\begin{eqnarray}
(\omega^{(1)}_{\rm edge})^2=\frac{\omega_{mag}^2+\omega_{em}^2}{2}+\sqrt{\left(\frac{\omega_{mag}^2+\omega_{em}^2}{2}\right)^2-\omega_H^2\omega_{em}^2} \nonumber \\
(k_y>0), \ \ \ \ \ \ \label{Edge_1}
\end{eqnarray}
\begin{eqnarray}
\omega^{(2)}_{\rm edge}=\omega^0_s\ \  (k_y<0),  \label{Edge_2}
\end{eqnarray}
with $\omega_{em}=ck_y/\sqrt{\epsilon}$ and $k_y>0$. 
The dispersion $\omega^{(1)}_{\rm edge}$ touches at $k_y=0$ the top of the $\omega_{-}$ branch of the bulk mode. 
The dispersion quadratically increases in small $k_y$ for $k_y\sim k^*$ due to the magnon, while it linearly increases for $k_y\gg k^*$ because the electromagnetic wave is dominant.  The other edge mode $\omega^{(2)}_{\rm edge}$ shows a flat 
dispersion as in Fig.~\ref{fig:Chern_Edge} (c) with $k_y<0$. 

An edge mode with a flat dispersion similar to Eq.~(\ref{Edge_2}) was also 
reported in a previous study of topological edge magnetoplasmon\cite{13486}. The magnetoplasmon 
is a coupled wave between the charge density and electric current density in a two-dimensional 
electron gas (2DEG) under a high magnetic field. 
The previous study\cite{13486} found two distinct edge modes in the 2DEG under the magnetic field, one edge 
mode with a flat dispersion and the other edge mode with a linear (chiral) dispersion. The edge mode with 
the flat dispersion carries only the electric current component, while the other edge mode carries 
both charge density and current components. Similarly to the topological magnetoedgeplasmon, the 
edge mode with the flat dispersion in the present system, Eq.~(\ref{Edge_2}), 
carries only the magnetization and the magnetic field components, 
but not the electric field component (see Appendix E). 
Meanwhile, the edge mode with the chiral dispersion, Eq.~(\ref{Edge_1}), is a 
coupled mode among magnetization, magnetic field and electric field (see Appendix E1). 

The edge mode with the flat dispersion in Eq.~(\ref{Edge_2}) can be regarded as 
the Damon-Eshbach surface mode in a ferromagnetic insulator slab with its surface being 
metalized\cite{1449610}. A dispersion of the surface 
mode of the surface-metalized ferromagnetic
insulator slab with a finite thickness $d$ exists only in the $k_y<0$ region in Fig.~\ref{fig:Chern_Edge} (d). When the thickness becomes much larger than the 
wavelength ($|k_y d|\gg 1$), the dispersion becomes flat when the exchange interaction is neglected\cite{1449610} and the 
saturated dispersion equals to Eq.~(\ref{Edge_2}). Note also that the boundary condition for the magnetic flux in the edge mode 
with the flat dispersion $\omega =\omega^0_s$ (see Appendix E) is the same as that in the 
surface-metalized ferromagnet, where the magnetic flux 
density along $x$ direction at the surface is zero due to the metalized surface\cite{1449610}.

\section{\label{sec:conclusion}Conclusion}
In this paper, we discuss the Berry curvature and topological edge modes that emerge 
from a hybridization between a magnon and an electromagnetic wave in a ferromagnetic insulator. 
By introducing a norm of eigenvector for the coupled wave based on the 
energy conservation, we reformulated the Landau-Lifshitz-Maxwell 
equation into a Hermitian eigenvalue equation. From the eigenvalue 
equation, we introduced the Berry curvature of the coupled waves between the magnon and 
the electromagnetic wave. When the wavevector of the coupled wave ${\bm k}$ is perpendicular to the magnetic 
field and magnetization, we found that the Berry curvature shows a prominent peak around a 
hybridization point between the magnon and the electromagnetic modes. The hybridization leads to two relevant hybrid 
modes; one is a magnon-like massive mode $(\omega=\omega_+)$ at ${\bm k}=0$  and the other is a photon-like massless mode $(\omega=\omega_-)$ at ${\bm k}=0$.  Around ${\bm k}\simeq 0$, the Berry curvature for the massless mode 
converges to zero, while that for the massive mode converges to a non-zero value. We found that the 
Chern number for the massive mode takes a non-zero integer ($\pm 1$), and consequently two chiral 
edge modes emerge inside the hybridization gap at a domain wall between two ferromagnetic insulators 
with opposite magnetizations. One of the two edge modes carries both a magnon and 
an electromagnetic wave, while the other edge mode is purely magnetic and can be regarded as 
the Damon-Eschbach surface chiral mode of the surface-metalized ferromagnetic insulator slab. 

Recently, the surface mode of the ferromagnet film in the dipole-exchange 
regime immune to backscattering is reported\cite{PhysRevLett.122.197201}. 
Our work provides an insight for the search of the chiral edge modes and stimulates future 
simulational and experimental studies on coupled waves between magnons and electromagnetic waves.

\begin{acknowledgments}
This work was supported by a MEXT KAKENHI Grant Number JP26100006, and by JST CREST Grant Number JPMJCR14F1. RS was 
supported by NBRP of China Grants No. 2014CB920901, No. 2015CB921104, and 
No. 2017A040215. 
\end{acknowledgments}

%%%%%%%%%%%%%%%%
\appendix

\section{Dispersion of the coupled wave between the magnon and the electromagnetic wave with the wavevector parallel to the magnetization}
\label{sec:}
%%%%%%%%%%%%%%%%%%%%%%%%%%%%%%

%%%%%%%%%%%%%%%%%%%%%%%%%%%%%%
In the main text, we consider the case with ${\bm k}\perp {\bm M}_0$. In this Appendix, we 
calculate 
the dispersion relation for the other case, the case with ${\bm k}\parallel  {\bm M}_0$. 
From Eq.~(\ref{dispersion relation}), the dispersion 
relation reduces to 
\begin{eqnarray}
&&\left[\omega \left(\omega^2-\omega_{em}^2\right)-\left(\omega_s\omega^2-\omega_H\omega_{em}^2\right)\right]\nonumber \\
&&\times \left[\omega \left(\omega^2-\omega_{em}^2\right)+\omega_s\omega^2-\omega_H\omega_{em}^2\right]=0. \label{dispersion relation_theta_0}
\end{eqnarray}
From this, we obtain the dispersion relations for three branches shown in Ref.~\onlinecite{stancil}.  Let $\omega_i$ ($i=1,2,3$) be the eigenfrequencies of the waves with $0< \!\ \omega_1<\omega_2<\omega_3$.
Among the three modes with positive frequencies, one is massive at $k=0$, 
$\omega(k=0)\not=0$, while the other two is massless at $k=0$. The dispersions of 
the massless modes take the following asymptotic forms 
around $k=0$, 
\begin{eqnarray}
\omega_1  \approx  \omega_{em}\sqrt{\frac{\omega_H}{\omega_s}}-\frac{\omega_M\omega_{em}^2}{2\omega_s^2}, \label{dispersion_theta_0_k_s_1}
\end{eqnarray} 
and
\begin{eqnarray}
\omega_2 \approx  \omega_{em}\sqrt{\frac{\omega_H}{\omega_s}}+\frac{\omega_M\omega_{em}^2}{2\omega_s^2}. \label{dispersion_theta_0_k_s_2}
\end{eqnarray} 
The dispersion of the massive mode has the following asymptotic form near $k=0$,
\begin{eqnarray}
\omega_3 \approx \omega_s +\frac{\omega_ M \omega _{em}^2}{ \omega_s^2}. \label{dispersion_theta_0_k_s_3}
\end{eqnarray}

\section{Calculation of the Berry curvature of the coupled wave between the magnon and the electromagnetic wave}
\label{sec:B}
%%%%%%%%%%%%%%%%%%%%%%%%%%%%%%

%%%%%%%%%%%%%%%%%%%%%%%%%%%%%%
In this Appendix, we give a detailed calculation of the 
Berry curvature for the coupled modes between 
a magnon and an electromagnetic wave for the case with 
${\bm k} \perp {\bm M}_0$. Let ${\bm M}_0$ to be along the $z$ axis. The two relevant branches with $\omega=\omega_{\pm}$ represent hybridized waves of $m_x$, $m_y$, $b_x$, $b_y$ and $e_z$. The eigenvalue equation 
for these five components, ${\bm x}_{\bm k}\equiv{}^t(m_{{\bm k},x},m_{{\bm k},y},b_{{\bm k},x},
b_{{\bm k},y},e_{{\bm k},z})$, is given by: 
\begin{eqnarray}
\tilde{H}^{\perp}_{\rm eff}(k,\varphi){\bm x}_{\bm k}&=
&\omega \gamma^{\perp}{\bm x}_{\bm k},
\end{eqnarray} 
where 
\begin{eqnarray}
&&\tilde{H}^{\perp}_{\rm eff}(k,\varphi)=\gamma ^{\perp}H_{\rm eff}(k,\varphi) \nonumber \\
&&=\scalebox{1.0}{$
\begin{pmatrix}
     0 & -i\frac{(4\pi)^2\omega_s^2}{\omega_M} & 0 & 4\pi i\omega_s & -4\pi ck_y \\
     i\frac{(4\pi)^2\omega_s^2}{\omega_M} & 0 & -4\pi i\omega_s & 0 & 4\pi ck_x \\
     0 & 4\pi i\omega_s & 0 & -i\omega_M & ck_y \\
     -4\pi i\omega_s & 0 & i\omega_M & 0 & -ck_x \\
     -4\pi ck_y & 4\pi ck_x & ck_y & -ck_x & 0 \\
    \end{pmatrix} 
$}. \nonumber \\  \label{Hamiltonian2}
\end{eqnarray} 
Here ${\bm k}\equiv k(\cos{\varphi},\ \sin{\varphi})$ is within the $xy$ plane, 
with $\varphi$ being the angle between the $x$ axis (see Fig.~\ref{fig:geometry_EMW}). The norm of 
the eigenvector is defined through the Hermitian matrix $\gamma^{\perp}$; 
\begin{eqnarray}
&&\gamma^{\perp}=\begin{pmatrix}
     \frac{(4\pi)^2\omega_s}{\omega_M} & 0 & -4\pi & 0 & 0 \\
     0 & \frac{(4\pi)^2\omega_s}{\omega_M} & 0 & -4\pi & 0 \\
     -4\pi & 0 & 1 & 0 & 0 \\
     0 & -4\pi & 0 & 1 & 0 \\
     0 & 0 & 0 & 0 & \epsilon \\
    \end{pmatrix}.
\end{eqnarray}

From the O(2) rotational symmetry in the Landau-Lifshitz-Maxwell equation, 
the eigenvector at finite $\varphi$ is related with that at $\varphi=0$ by the O(2) 
transformation, 
\begin{eqnarray}
&&
\tilde{H}^{\perp}_{\rm eff}(k,\varphi =0)
\tilde{\bm{x}}_k={\omega}\gamma ^{\perp}\tilde{\bm{x}}_k,
\label{eq:phi0}\\
&&\bm{x}_{\bm{k}}
=\begin{pmatrix}
{\bm U}_2(\varphi) &  &  \\
 & {\bm U}_2(\varphi) &  \\
 &  & 1 \\
\end{pmatrix}
\tilde{\bm{x}}_{k}, \label{EV_GT}
\end{eqnarray} 
where
\begin{eqnarray}
{\bm U}_2(\varphi)=
\begin{pmatrix}
\cos\varphi & -\sin\varphi \\
\sin\varphi & \cos\varphi \\
\end{pmatrix}. \label{GT}
\end{eqnarray} 
The dependence on $\varphi$ and $k$ in ${\bm x}_{\bm k}$ 
is now factorized into ${\bm U}_2(\varphi)$ and 
$\tilde{\bm x}_{k}$. $\tilde{\bm x}_{k}$ is an  
eigenstate of the following Hermitian matrix,  
\begin{eqnarray}
&&\tilde{H}^{\perp}_{\rm eff}(k,\varphi =0) \nonumber \\
&&=\begin{pmatrix}
     0 & -i\frac{(4\pi)^2\omega_s^2}{\omega_M} & 0 & 4\pi i\omega_s & 0 \\
     i\frac{(4\pi)^2\omega_s^2}{\omega_M} & 0 & -4\pi i\omega_s & 0 & 4\pi ck \\
     0 & 4\pi i\omega_s & 0 & -i\omega_M & 0 \\
     -4\pi i\omega_s & 0 & i\omega_M & 0 & -ck \\
     0 & 4\pi ck & 0 & -ck & 0 \\
    \end{pmatrix} . \nonumber \\
\end{eqnarray}

By using the factorized form for ${\bm x}_{\bm k}$
, the Berry curvature is calculated as
\begin{eqnarray}
\Omega _{z, n}(\bm{k})&=&i\epsilon _{\alpha \beta}\frac{\partial \bm{x}^\dagger_{\bm{k}}}{\partial k_\alpha }\gamma^{\perp}\frac{\partial \bm{x}_{\bm{k}}}{\partial k_\beta  }\nonumber\\
&=&\frac{1}{k}\frac{\partial}{\partial k}
\Big(\tilde{\bm{x}}^\dagger_k\Gamma \tilde{\bm{x}}_k \Big), 
\label{omeganz} \\ 
\Gamma &=& \begin{pmatrix}
  0 & -\frac{(4\pi)^2\omega_s}{\omega_M} & 0 & 4\pi  & 0 \\
  \frac{(4\pi)^2\omega_s}{\omega_M} & 0 & -4\pi & 0 & 0 \\
  0 & 4\pi & 0 & -1 & 0 \\
  -4\pi & 0 & 1 & 0 & 0 \\
  0 & 0 & 0 & 0 & 0 
\end{pmatrix},
\end{eqnarray}
with the normalization condition $\tilde{\bm{x}}^\dagger_k\gamma^{\perp}\tilde{\bm{x}}_k=1$. 
To evaluate Eq.~(\ref{omeganz}), it is convenient to introduce 
an unnormalized eigenstate $\tilde{\bm{X}}_k$, which is related with $\tilde{\bm{x}}_k$ by
\begin{align}
\tilde{\bm{x}}_k \equiv \frac{ \tilde{\bm{X}}_k  }
{\sqrt{\tilde{\bm{X}}^\dagger_k \gamma^{\perp}\tilde{\bm{X}}_k}}. 
\end{align}
In terms of the unnormalized eigenstate, the Berry curvature is given by
\begin{eqnarray}
&&\Omega _{z, n}(\bm{k})=\frac{1}{k}\frac{\partial}{\partial k}
\left(\frac{\tilde{\bm{X}}_k\Gamma \tilde{\bm{X}}_k}{
\tilde{\bm{X}}_k\gamma^{\perp}\tilde{\bm{X}}_k}
\right).
\label{eqn:Xk}
\end{eqnarray}
From the Hermitian eigenvalue equation, the eigenstate satisfies
\begin{eqnarray}
\begin{pmatrix}
      -\frac{(4\pi)^2\omega_s\omega}{\omega_M} & -i\frac{(4\pi)^2\omega_s^2}{\omega_M} & 4\pi\omega & 4\pi i\omega_s & 0 \\
     i\frac{(4\pi)^2\omega_s^2}{\omega_M} & -\frac{(4\pi)^2\omega_s\omega}{\omega_M}  & -4\pi i\omega_s & 4\pi\omega & 4\pi ck \\
     4\pi\omega & 4\pi i\omega_s & -\omega & -i\omega_M & 0 \\
     -4\pi i\omega_s & 4\pi\omega & i\omega_M & -\omega & -ck \\
     0 & 4\pi ck & 0 & -ck & -\epsilon \omega \\
    \end{pmatrix}
\tilde{\bm{X}}_k={\bm 0}. \label{Hermiation_eigenvalue_equation_EMW} \nonumber \\
\end{eqnarray}
Thus, we have 
\begin{eqnarray}
\tilde{\bm{X}}_k=\begin{pmatrix}
 -i\frac{\omega_M ck}{4\pi(\omega^2-\omega_s^2)} \\
  \frac{\omega_s\omega_M ck}{4\pi\omega(\omega^2-\omega_s^2)} \\
 0 \\
 -\frac{ck}{\omega} \\
 1 
\end{pmatrix}. \label{Xk-def}
\end{eqnarray}
By using this wavefuntion and Eq.~(\ref{eqn:Xk}), we obtain the Berry curvature for the two eigenmodes with $\omega=\omega_{\pm}$:
\begin{eqnarray}
\Omega_{z,\pm}(k)=\frac{1}{k}\frac{\partial}{\partial k}\left(\frac{\frac{\omega_{\pm}}{\omega_s}(\omega_{\pm}^2-\omega_{em}^2)}{(2\omega_{\pm}^2-\omega_s^2-\omega_{em}^2)}\right). \label{BC_Appendix}
\end{eqnarray}

\section{Wavenumber and Berry curvature at the crossing point of the dispersions for the electromagnetic wave and the magnon}
\label{sec:C}
Here we present a detailed calculation of the peak value of the Berry curvature at the 
crossing point between magnon and photon mode. The wavevector at the crossing point $k^{*}$ 
is defined by
%\begin{eqnarray}
%&& \frac{c^2}{\epsilon }{k^\ast}^2=(\omega_H+\omega_M)^2,
%\end{eqnarray} 
\begin{eqnarray}
k^*= \sqrt{\frac{\epsilon}{c^2}}(\omega_H+\omega_M).
\end{eqnarray}  
The dispersion and the Berry curvature around $k=k^{*}$ are given by 
\begin{align}
{\omega}_{\pm} \simeq & \sqrt{\frac{c^2(k^2+{k^*}^2)}{2\epsilon }} \nonumber \\
& \hspace{-0.7cm} 
\times\left(1\pm\frac{1}{(k^2+{k^*}^2)}\sqrt{\frac{(k^2-{k^*}^2)^2}{4}+\left(\frac{\epsilon \zeta k}{c^2}\right)^2}\right). \label{Dis_CP}
\end{align}
Using Eqs.~(\ref{BC_Appendix}, \ref{Dis_CP}), we obtain peak values of the Berry curvature;
\begin{eqnarray}
\Omega_{z,\pm}(k=k^*)&=&\mp \frac{\tilde {\omega}^2}{2{k^*}^3\zeta}=\mp \frac{1}{{2k^*}^2\Delta \omega/\tilde{\omega}}.
\end{eqnarray}

\section{Hermitian eigenvalue problems in the other bases}  
In the main text, we use the basis for the eigenvector as 
${\bm x}_{{\bm k},1} \equiv 
({\bm m}_{\bm k}, {\bm b}_{\bm k}, {\bm e}_{\bm k})^t$. 
Using ${\bm b}_{\bm k}= {\bm h}_{\bm k}+4\pi{\bm m}_{\bm k}$, one can change the 
basis for the eigenvector into either 
${\bm x}_{{\bm k},2 } \equiv ({\bm m}_{\bm k}, {\bm h}_{\bm k}, {\bm e}_{\bm k})^t$ 
or ${\bm x}_{{\bm k},3 } \equiv ({\bm b}_{\bm k}, {\bm h}_{\bm k}, {\bm e}_{\bm k})^t$.  
In terms of ${\bm x}_{{\bm k},2}$, the Hermitian Hamiltonian in the 
eigenvalume problem changes into 
\begin{eqnarray}
H_2({\bm k})&=&\scalebox{1.0}{$
\begin{pmatrix}
     0 & -i\frac{(4\pi)^2\omega_H^2}{\omega_M} & 0 & 4\pi i\omega_H & 0 \\
     i\frac{(4\pi)^2\omega_H^2}{\omega_M} & 0 & -4\pi i\omega_H & 0 & 0 \\
     0 & 4\pi i\omega_H & 0 & -i\omega_M & ck_y \\
     -4\pi i\omega_H & 0 & i\omega_M & 0 & -ck_x \\
     0 & 0 & ck_y & -ck_x & 0 \\
    \end{pmatrix} 
$}. \nonumber \\  \label{Hamiltonian3}
\end{eqnarray} 
In the new basis, a norm of the eigenvectors should be redefined as  
$\left< {\bm x}_{{\bm k},2}\right|\gamma_{2} \left|{\bm x}_{{\bm k},2} \right>$ 
with 
\begin{eqnarray}
&&\gamma_2=\begin{pmatrix}
     \frac{(4\pi)^2\omega_H}{\omega_M} & 0 & 0 & 0 & 0 \\
     0 & \frac{(4\pi)^2\omega_H}{\omega_M} & 0 & 0 & 0 \\
     0 & 0 & 1 & 0 & 0 \\
     0 & 0 & 0 & 1 & 0 \\
     0 & 0 & 0 & 0 & \epsilon \\
    \end{pmatrix}.
\end{eqnarray}
(compare this with Eqs.~(\ref{ut}), (\ref{ued}), and (\ref{umd}). We can also choose another basis, 
${\bm x}_{{\bm k},3 }\equiv ({\bm b}_{\bm k}, {\bm h}_{\bm k}, {\bm e}_{\bm k})^t$, with 
\begin{eqnarray}
H_3({\bm k})&=&\scalebox{1.0}{$
\begin{pmatrix}
     0 & -i\frac{\omega_H^2}{\omega_M} & 0 & i\frac{\omega_H\omega_s}{\omega_M} & 0 \\
     i\frac{\omega_H^2}{\omega_M} & 0 & -i\frac{\omega_H\omega_s}{\omega_M} & 0 & 0 \\
     0 &  i\frac{\omega_H\omega_s}{\omega_M} & 0 & -i\frac{\omega_s^2}{\omega_M} & ck_y \\
     - i\frac{\omega_H\omega_s}{\omega_M} & 0 & i\frac{\omega_s^2}{\omega_M} & 0 & -ck_x \\
     0 & 0 & ck_y & -ck_x & 0 \\
    \end{pmatrix} 
$}, \nonumber \\  \label{Hamiltonian4}
\end{eqnarray} 
and 
\begin{eqnarray}
&&\gamma_3=\begin{pmatrix}
     \frac{\omega_H}{\omega_M} & 0 & -\frac{\omega_H}{\omega_M} & 0 & 0 \\
     0 & \frac{\omega_H}{\omega_M} & 0 & -\frac{\omega_H}{\omega_M} & 0 \\
     -\frac{\omega_H}{\omega_M} & 0 & \frac{\omega_s}{\omega_M} & 0 & 0 \\
     0 & -\frac{\omega_H}{\omega_M} & 0 & \frac{\omega_s}{\omega_M} & 0 \\
     0 & 0 & 0 & 0 & \epsilon \\
    \end{pmatrix}.
\end{eqnarray}
The norm in this basis is defined as $\left< {\bm x}_{{\bm k},3}\right|\gamma_{3} \left|{\bm x}_{{\bm k},3} \right>$. In accordance with the change of the norm, the Berry curvature in these 
new bases are defined by Eq.~(\ref{berrycurvature}) with a replacement of 
$\gamma^{\perp}$ and ${\bm x}_{{\bm k}}$ by $\gamma_2$ and ${\bm x}_{{\bm k},2}$ 
or by $\gamma_3$ and ${\bm x}_{{\bm k},3}$ respectively. 
It is important to note that these different 
formulae give the same calculation result of the Berry curvature as Eq.~(\ref{BC_MEMW}). 

\section{Calculation of the edge-mode solutions of the coupled wave between the magnon 
and the electromagentic wave}
In the main text, we discuss the emergence of the chiral edge modes 
at the boundary between the 
two ferromagnetic regions with an opposite magnetization and magnetic field. 
In the following, we will give a detailed derivation of the edge modes and 
their dispersions. From Eq.~(\ref{kperpM}) with the replacement of $k_x$ by $\pm i\kappa\ (\kappa>0)$, the eigenfrequencies of the edge modes shold satisfy the following equation;
\begin{align}
\omega^4 - (\tilde{\omega}^2_{em}+\omega^2_s) &\omega^2 + \tilde{\omega}^2_{em}\omega^2_{mag} = 0,  
\label{dispersion-for-edge-a} \\
\tilde{\omega}^2_{em} &\equiv \frac{c^2 (k_y^2 - \kappa^2)}{\epsilon}. \label{dispersion-for-edge-b}
\end{align}  
Unnormalized eigenvectors for the edge modes are obtained from 
Eqs.~(\ref{Xk-def}) and (\ref{EV_GT}) with the replacement of 
$k_x$ by $+i\kappa\ (\kappa>0)$ for the region II ($x>0$) and 
by $-i\kappa$ for the region I ($x<0$), Eqs.~(\ref{Xk-def}) and (\ref{EV_GT}) 
give the unnormalized eigenvectors at the both sides of the 
boundary, 
\begin{eqnarray}
& &\psi_{k_y}(x,y)=\nonumber \\
& &
\begin{pmatrix}
m_{x,k_y}  \\
m_{y,k_y}  \\
b_{x,k_y}  \\
b_{y,k_y}  \\
e_{z,k_y} \\
\end{pmatrix}
= C_{\pm}\begin{pmatrix} 
 \pm \frac{\omega_M \!\ c\kappa}{4\pi} -\frac{\omega_s\omega_M ck_y}{4\pi\omega}\\
  - i\frac{\omega_M  \!\ ck_y}{4\pi}\pm i\frac{\omega_s\omega_M c \kappa}{4\pi\omega} \\
 \frac{ck_y}{\omega}(\omega^2-\omega_s^2) \\
 \mp i\frac{c\kappa}{\omega}(\omega^2-\omega_s^2) \\ 
 (\omega^2-\omega_s^2)
\end{pmatrix} e^{\mp \kappa x+ik_yy}, \nonumber \\ && \ \ \ \ \ \ \ \ \ \ \ \ \ \ \ \ \ \ \ \ \ \ \ \ \ \ \   \ \ \ \ \ \ \ \ \ \   \ \ \ \ \ \ \ \ \ \    (x>0\ (x<0)) \label{connection}
\end{eqnarray}
where $C_{\pm}$ are constants.
Here, we note that $\omega_H=\omega^0_H\equiv gH_0$, $\omega_M=\omega^0_M\equiv 4\pi gM_0$, and $\omega_s=\omega^0_s\equiv\omega^0_H+\omega^0_M$ for $x>0$ and $\omega_H=-\omega^0_H$, $\omega_M=-\omega^0_M$, and $\omega_s=-\omega^0_s$ for $x<0$.
Next, we need to determine the constant factors $C_{\pm}$ so as to satisfy the appropriate boundary conditions:
\begin{eqnarray}
b_x(x=0+)&=&b_x(x=0-),\\
e_z(x=0+)&=&e_z(x=0-),\\
h_y(x=0+)&=&h_y(x=0-).
\end{eqnarray}  
In the following, to calculate edge-modes solutions, we study cases the $\omega\neq \omega^0_s $ and $\omega= \omega^0_s$ separately. 

\subsection{edge-mode solution with $\omega\neq \omega^0_s$}
Let us first consider an edge-mode solution with $\omega\neq \omega^0_s$.
From Eq.~(\ref{connection}), to satisfy the boundary conditions 
for $b_x$ and $e_z$ at $x=0$ we need to set $C_+=C_-$. 
Then the boundary condition for $h_y$ at $x=0$ gives a relation 
between $\kappa$ and $k_y$,
\begin{align}
\kappa = \frac{\omega^0_{M} \!\ \omega}{\omega^2 -\omega^2_{mag}} k_y. \label{relation}
\end{align}
A substitution of Eq.~(\ref{relation}) into 
Eqs.~(\ref{dispersion-for-edge-a}) and (\ref{dispersion-for-edge-b}) leads to  
the dispersion relation between $k_y$ and $\omega$ for localized modes: 
\begin{align}
\omega^4 - (\omega^2_{mag} + \omega^2_{em})\omega^2 + \omega^2_H 
\omega^2_{em}= 0, \label{dispersion-for-edge-2}
\end{align}
with $\omega_{em} \equiv ck_y/\sqrt{\epsilon}$ for $\omega \neq \omega_{mag}$. It gives the chiral dispersion, Eq.~(\ref{Edge_1}), 
where $k_y>0$ is required by the positiveness of $\kappa$.

The edge mode with the chiral dispersion involves both 
a magnetization and an electric field. For $k_y\rightarrow 0$, the chiral edge mode becomes 
magnonic, 
\begin{align}
\begin{pmatrix}
m_{x,k_y} (x,y) \\
m_{y,k_y} (x,y) \\
b_{x,k_y} (x,y)\\
b_{y,k_y} (x,y) \\
e_{z,k_y} (x,y) \\
\end{pmatrix} =C_{\pm}\begin{pmatrix}
 \frac{ 1}{4\pi \sqrt{2}}\sqrt{\frac{\omega_{M}}{\omega_s}}  \\
  \mp\frac{i}{4\pi\sqrt{2}}\sqrt{\frac{\omega_M}{\omega_H}} \\
 0 \\
  \mp\frac{i}{\sqrt{2}}\sqrt{\frac{\omega_M}{\omega_H}} \\
 0 
\end{pmatrix} \!\ e^{\mp \kappa x +ik_y y}, \nonumber \\  (x>0\ (x<0)). \label{magnon-like}
\end{align}
For $k_y\gg k^{*}$, the chiral mode becomes photonic, 
\begin{align}
\begin{pmatrix}
m_{x,k_y} (x,y) \\
m_{y,k_y} (x,y) \\
b_{x,k_y} (x,y)\\
b_{y,k_y} (x,y) \\
e_{z,k_y} (x,y) \\
\end{pmatrix} =C_{\pm}\begin{pmatrix}
 0 \\
 0 \\
 1/\sqrt{2} \\
 0 \\
 1/\sqrt{2 \epsilon} 
\end{pmatrix} \!\  e^{\mp \kappa x +ik_y y}, \nonumber \\  (x>0\ (x<0)). \label{photon-like}
\end{align}

\subsection{edge-mode solution with $\omega=\omega^0_s$}
First, Eq.~(\ref{connection}) satisfies the boundary conditions for $b_x$ and $e_z$, while the boundary condition for $h_y$ is satisfied by setting $C_-=-C_+$. Then, by combining $\omega = \omega^0_s$,
we obtain with Eqs.~(\ref{dispersion-for-edge-a}) and (\ref{dispersion-for-edge-b}) 
relates $k_y$ with $\kappa$ as,
\begin{align}
\tilde{\omega}^2_{em} \equiv \frac{c^2 (k_y^2-\kappa^2)}{\epsilon} = 0,\ \  \Rightarrow k_y=\pm \kappa
\label{anti2}
\end{align}
A case with $k_y= +\kappa \!\ (>0)$ makes all the components in 
Eq.~(\ref{connection}) to be zero, giving no physical solution. The other case with 
$k_y=-\kappa \!\ (<0)$ gives a physical edge-mode solution with the flat dispersion, 
Eq.~(\ref{Edge_2}). From Eq.~(\ref{connection}), the edge mode with the flat 
dispersion involves only the magnetization and the magnetic field;
\begin{align}
\begin{pmatrix}
m_{x,k_y} (x,y) \\
m_{y,k_y} (x,y) \\
b_{x,k_y} (x,y)\\
b_{y,k_y} (x,y) \\
e_{z,k_y} (x,y) \\
\end{pmatrix} = C_{\pm}\begin{pmatrix}
 \pm\frac{1}{4\sqrt{2}\pi}\sqrt{\frac{\omega_M}{\omega_s}} \\
 i\frac{1}{4\sqrt{2}\pi}\sqrt{\frac{\omega_M}{\omega_s}} \\
 0 \\
 0 \\
 0 
\end{pmatrix} e^{\mp \kappa x +ik_y y},  \nonumber \\  (x>0\ (x<0)). \label{flat-band-eigenvector}  
\end{align}


\nocite{*}

%\bibliography{berrycurvature_magnon_electromagnetic}% Produces the bibliography via BibTeX.

%

\end{document}